\documentclass[onecolumn,showpacs]{revtex4}

\usepackage{graphicx}
\usepackage{dcolumn}
\usepackage{bm}

\input epsf

\begin{document}

\title{\Large Quasi-spherical collapse with cosmological constant}

\author{\bf Ujjal Debnath$^1$\footnote{ujjaldebnath@yahoo.com}, Soma Nath$^2$ and ~Subenoy
Chakraborty$^2
$\footnote{subenoyc@yahoo.co.in}}
\affiliation{$^1$Department of Mathematics, Bengal Engineering and
Science University,\\ Shibpur, Howrah-711 103, India.\\
$^2$Department of Mathematics, Jadavpur University, Calcutta-32,
India.}

\date{\today}

\begin{abstract}
The junction conditions between static and non-static space-times
are studied for analyzing gravitational collapse in the presence
of a cosmological constant. We have discussed about the apparent
horizon and their physical significance. We also show the effect
of cosmological constant in the collapse and it has been shown
that cosmological constant slows down the collapse of matter.
\end{abstract}


\maketitle

\section{\normalsize\bf{Introduction}}

Gravitational collapse is one of the most important problem in
classical general relativity. Usually, the formation of compact
stellar objects such as white dwarf and neutron star are preceded
by a period of collapse. Hence for astrophysical collapse, it is
necessary to describe the appropriate geometry of interior and
exterior regions and to determine proper junction conditions
which allow the matching of these regions.\\

 The study of gravitational collapse was started by
Oppenheimer and Snyder [1]. They studied  collapse of dust with a
static Schwarzschild exterior while interior space-time is
represented by Friedman like solution. Since then several authors
have extended the above study of collapse of which important and
realistic generalizations are the following: (i) the static
exterior was studied by Misner and Sharp [2] for a perfect fluid
in the interior, (ii) using the idea of outgoing radiation of the
collapsing body by Vaidya [3], Santos and collaborations [4-9]
included the dissipation in the source by allowing radial heat
flow (while the body undergoes radiating collapse). Ghosh and
Deskar [10] have considered collapse of a radiating star with a
plane symmetric boundary (which has a close resemblance with
spherical symmetry [11]) and have concluded with some general
remarks. On the othehand, Cissoko et al [12] have studied junction
conditions between static and non-static space-times for analyzing
gravitational collapse in the presence of dark energy has been
investigated by Mota et al [13] and Cai et al [14]. The effect
of cosmological constant (a source of dark energy) in cosmology
has been shown by Lahav et al [15] and Antolinez et al [16].\\

So far most of the studies have considered in a star whose
interior geometry is spherical. But in the real astrophysical
situation the geometry of the interior of a star may not be
exactly spherical, rather quasi-spherical in form. Recently,
solutions for arbitrary dimensional Szekeres' model with perfect
fluid (or dust) [17] has been found for quasi-spherical or
quasi-cylindrical symmetry of the space-time. Also a detailed
analysis of the gravitational collapse [18, 19] has been done for
quasi-spherical symmetry of the Szekeres' model. It has also been
studied junction conditions between quasi-spherical interior
geometry of radiating star and exterior Vaidya metric [20]. In
this paper, we have considered the interior space-time $V^{-}$ by
Szekeres' model [17, 21] while for exterior geometry $V^{+}$ we
have considered Schwarzschild-de-Sitter space-time. The plan of
the paper is as follows: The junction conditions has been
presented in section II. The apparent horizons and their physical
interpretations are shown in section III. The paper ends with a
short conclusion in section IV.

\section{\normalsize\bf{Junction conditions}}

Let us consider a time-like $3 D$ hypersurface $\Sigma$, which
divides $4 D$ space-time into two distinct $4 D$ manifolds $V^{-}$
and $V^{+}$. For junction conditions we follow the modified
version of Israel [22] by Santos [4, 5]. Now the geometry of the
space-time $V^{-}$ inside the boundary $\Sigma$ is given by the
Szekeres space-time

\begin{equation}
ds_{-}^{2}=-dt^{2}+e^{2\alpha}dr^{2}+e^{2\beta}(dx^{2}+dy^{2})
\end{equation}

where $\alpha$ and $\beta$ are functions of all space-time
variables.\\

The metric co-efficients $\alpha$ and $\beta$ have the explicit
form for dust matter with cosmological constant $\Lambda$ [13, 16]

\begin{equation}
e^{\beta}=R(t,r)e^{\nu(r,x,y)}
\end{equation}
\begin{equation}
e^{\alpha}=R'+R\nu'
\end{equation}

The evolution equation for $R$ is

\begin{equation}
\dot{R}^{2}=f(r)+\frac{F(r)}{R}+\frac{\Lambda}{3}~R^{2}
\end{equation}

and $\nu$ has the explicit form

\begin{equation}
e^{-\nu}=A(r)(x^{2}+y^{2})+B_{1}(r)x+B_{2}(r)y+C(r)
\end{equation}

where $F(r)$ ($>0$) and $f(r)$ are arbitrary functions of $r$ and
$A(r),~B_{1}(r),~B_{2}(r)$ and $C(r)$ are arbitrary functions of
$r$ along with the restriction

\begin{equation}
B_{1}^{2}+B_{2}^{2}-4AC=f(r)-1
\end{equation}

Assuming $x=cot(\theta/2)cos\phi,~y=cot(\theta/2)sin\phi$,
equation (1) becomes

\begin{equation}
ds_{-}^{2}=-dt^{2}+e^{2\alpha}dr^{2}+\frac{1}{4}~e^{4\beta}cosec^{4}(\theta/2)(d\theta^{2}+sin^{2}\theta
d\phi^{2})
\end{equation}

For exterior space-time $V^{+}$ to $\Sigma$, we have considered
the Schaezschild-de-Sitter space-time

\begin{equation}
ds_{+}^{2}=-N(z)dT^{2}+\frac{1}{N(z)}dz^{2}+(d\theta^{2}+sin^{2}\theta
d\phi^{2})
\end{equation}

where $N(z)=\left(1-\frac{2M}{z}-\frac{\Lambda}{3}~z^{2}\right)$,
$M$ is a constant.\\

The intrinsic metric on the boundary $\Sigma$ of the hypersurface
$r=r_{\Sigma}$ is given by

\begin{equation}
ds_{\Sigma}^{2}=-d\tau^{2}+A^{2}(\tau)(d\theta^{2}+sin^{2}\theta
d\phi^{2})
\end{equation}

 Now Israel's junction conditions (as described
by Santos [4, 5]) are\\

(i)~ The continuity of the line element i.e.,
\begin{equation}
(ds^{2}_{-})_{\Sigma}=(ds^{2}_{+})_{\Sigma}=ds^{2}_{\Sigma}
\end{equation}

where $(~ )_{\Sigma}$ means the value of (~ ) on $\Sigma$.\\

(ii)~ The continuity of extrinsic curvature over $\Sigma$ gives
\begin{equation}
[K_{ij}]=K_{ij} ^{+}-K_{ij}^{-}=0~,
\end{equation}
where due to Eisenhart [23] the extrinsic curvature has the
expresion

\begin{equation}
K_{ij}^{\pm}=-n_{\sigma}^{\pm}\frac{\partial^{2}\chi^{\sigma}_{\pm}}{\partial\xi^{i}
\partial \xi^{j} }-n^{\pm}_{\sigma}\Gamma^{n}_{\mu
\nu}\frac{\partial\chi^{\mu}_{\pm}}{\partial\xi^{i}}\frac{\partial\xi^{\nu}_{\pm}}
{\partial\xi^{j}}
\end{equation}
Here $ \xi^{i}=(\tau,x,y)$ are the intrinsic co-ordinates to
$\Sigma, ~\chi^{\sigma}_{\pm},~ \sigma =0,1,2,3$ are the
co-ordinates in $V^{\pm}$ and $n_{\alpha}^{\pm}$ are the
components of the normal vector to $\Sigma$ in the co-ordinates
$\chi^{\sigma}_{\pm}$.\\

Now for the interior space-time described by the metric (1) the
boundary of the interior matter distribution (i.e., the surface
$\Sigma$) is characterized by
\begin{equation}
f(r,t)=r-r_{_{\Sigma}}=0
\end{equation}
where $r_{_{\Sigma}}$ is a constant. As the vector with components
$\frac{\partial f}{\partial \chi^{\sigma}_{-} }$ is orthogonal to
$\Sigma$ so we take
$$n_{\mu}^{-}=(0,e^{\alpha},0,0).$$
So comparing the metric ansatzs given by equations (1) and (9) for
$dr=0$ we have from the continuity relation (10)

\begin{equation}
\frac{dt}{d\tau}=1,~~A(\tau)=e^{\beta}~~~~
\text{on}~~~r=r_{_{\Sigma}}
\end{equation}

Also the components of the extrinsic curvature for the interior
space-time are

\begin{equation}
K^{-}_{\tau\tau}=0~~~\text{and}~~~  K
^{-}_{\theta\theta}=cosec^{2}\theta K
^{-}_{\phi\phi}=\left[\frac{1}{4}~\beta'
e^{2\beta-\alpha}cosec^{4}(\theta/2)\right]_{\Sigma}.
\end{equation}

On the other hand for the exterior Schwarzschild-de-Sitter metric
described by the equation (8) with its interior boundary, given by
\begin{equation}
f(z,T)=z-z_{_{\Sigma}}(T) =0
\end{equation}

the unit normal vector to $\Sigma$ is given by

\begin{equation}
n_{\mu}^{+}=\left(N-\frac{1}{N}\left(\frac{dz}{dT}\right)^{2}\right)^{-1/2}\left(-
\frac{dz}{dT},1,0,0\right)
\end{equation}

and the components of the extrinsic curvature are

\begin{equation}
K^{+}_{\tau\tau}=\left[\frac{N\ddot{T}}{z}+\frac{dN}{dz}~\dot{T}\right]_{\Sigma}
\end{equation}
and
\begin{equation}
K^{+}_{\theta\theta}=cosec^{2}\theta
K^{+}_{\phi\phi}=\left[\dot{T}Nz \right]_{\Sigma}
\end{equation}

Hence the continuity of the extrinsic curvature due to junction
condition (eq. (11)) gives

\begin{equation}
N=
\left[\frac{1}{4}~e^{2\beta}~cosec^{4}(\theta/2)\left(e^{-2\alpha}\beta'^{2}-
\dot{\beta}^{2}\right)\right]_{\Sigma}
\end{equation}
and
\begin{equation}
\dot{T}_{\Sigma}=\left[2sin^{2}(\theta/2)\left(e^{\beta-\alpha}\beta'^{2}-
e^{\beta+\alpha}\dot{\beta}^{2}\right)^{-1}\right]_{\Sigma}
\end{equation}

Now using the junction condition (20) with the help of equations
(2), (3) and (4), we have (on the boundary) [18]

\begin{equation}
\frac{1}{2}\dot{R}^{2}-\frac{M}{R}-\frac{\Lambda}{6}~R^{2}=0
\end{equation}

which can be interpreted as the energy conservation equation on
the boundary. It is to be noted that the cosmological term leads
to a repulsive term to the Newtonian potential [16] i.e.,

\begin{equation}
\phi(R)=\frac{M}{R}+\frac{\Lambda}{6}~R^{2}
\end{equation}

\section{\normalsize\bf{Trapped Surfaces : Cosmological and Black Hole Horizons}}

As the present space-time geometry is complicated, so it is
difficult to find the formation of event horizon. However, trapped
surfaces which are space-like 2-surfaces with normals on both
sides are future pointing converging null geodesic families, may
be considered here. In fact, if the 2-surface $S_{r,t}$
($r=$constant, $t=$constant) is a trapped surface then it and its
entire future development lie behind the event horizon unless the
density falls off fast enough at infinity. So if $K^{\mu}$ is the
tangent vector field to the null geodesics orthogonal to the
trapped surface then $K^{\mu}$ should satisfy (i)
$K_{\mu}K^{\mu}=0$, (ii) $K^{\mu}_{;\nu}K^{\nu}=0$.\\

Also the convergence (or divergence) of the null geodesics on the
trapped surface is characterized by the sign of the scalar
$K^{\mu}_{;\mu}$ ($K^{\mu}_{;\mu}<0$ for convergence,
$K^{\mu}_{;\mu}>0$ for divergence). It is to be noted that the
inward geodesics converges initially and throughout the collapsing
process but the outward geodesics diverges initially but becomes
convergent after a time $t_{ah}(r)$ (the time of formation of
apparent horizon) given by

\begin{equation}
\dot{R}^{2}=1+f(r)
\end{equation}

Then from the evolution equation (4), we have

\begin{equation}
\Lambda R^{3}-3R+3F(r)=0
\end{equation}

The possible solutions of equation (25) for different choices of
$\Lambda$ and $F(r)$ are shown in the TABLE.\\

For marginally bound case (i.e., $f(r)=0$) the evolution equation
(4) can be solved as

\begin{equation}
t_{c}(r)-t=\frac{2}{\sqrt{3\Lambda}}~sinh^{-1}\left[\sqrt{\frac{\Lambda}{3F}}~R^{3/2}
\right]
\end{equation}

where, $t=t_{c}(r)$ is the time of collapse of a shell of radius
$r$ (i.e., $R=0$ at $t=t_{c}(r)$).\\

\newpage

\begin{center}
{\bf \text TABLE} \\\vspace{.5cm}
\begin{tabular}{|c|c|} \hline\hline
~~~&~~\\
~ {\bf Restrictions on $\Lambda,~F(r)$}~~~&~~{\bf Solutions of
eq.(25):
~Different horizons}\\
~~~&~~\\
 \hline\hline

~~~&~~\\
(a)~$\Lambda=0$~~~&~~$R=F(r)$, Schwarzschild horizon\\
~~~&~~\\
 \hline

~~~&~~\\
(c)~$F(r)=0$~~~&~~$R=0$ ~~(black hole)\\
                  ~~~~~&~~$R=\pm \frac{1}{\sqrt{\Lambda}}$~~~(de-Sitter horizon)\\
~~~&~~\\
 \hline

~~~&~~\\
(d)~$F(r)<\frac{2}{3}~\frac{1}{\sqrt{\Lambda}}$~~~&~~Two horizons:\\
~~~&~~\\
   ~~~~~~&~~$R_{1}=\frac{2}{\sqrt{\Lambda}}~cos(\theta/3)$\\
   ~~~~~~&~~$R_{2}=\frac{1}{\sqrt{\Lambda}}~[-cos(\theta/3)+\sqrt{3}~sin(\theta/3)]$\\
   ~~~~~~&~~$cos\theta=-\frac{3}{2}~\sqrt{\Lambda~F(r)}$\\
   ~~~~~~&~~$0\le R_{2}\le \sqrt{\Lambda}\le R_{1}\le \frac{\sqrt{3}}{\sqrt{\Lambda}}$\\
~~~&~~\\
 \hline

~~~&~~\\
(e)~$F(r)=\frac{2}{3}~\frac{1}{\sqrt{\Lambda}}$~~~&~~$R=0$ ~~$R=\frac{1}{\sqrt{\Lambda}}$\\
~~~&~~\\
 \hline

~~~&~~\\
(f)~$F(r)>\frac{2}{3}~\frac{1}{\sqrt{\Lambda}}$~~~&~~no horizon\\
~~~&~~\\
 \hline

\end{tabular}
\end{center}

\vspace{.5cm}

Hence the time of formation of apparent horizon $t_{ah}(r)$ is
given by

\begin{equation}
t_{ah}(r)=t_{c}(r)-\frac{2}{\sqrt{3\Lambda}}~sinh^{-1}\left[\sqrt{\frac{\Lambda}{3F}}~R_{H}^{3/2}
\right]
\end{equation}

where $R_{H}$ is a root of the equation (25).\\

Thus from the above table we see that in the fourth case (i.e.,
$F(r)<\frac{2}{3}~\frac{1}{\sqrt{\Lambda}}$) we have two horizons
namely cosmological and black hole horizons ($R_{1}\ge R_{2}$)
and let $t_{1}$ and $t_{2}$ be their time of formation then from
equation (27), $t_{1}\le t_{2}$, i.e., cosmological horizon forms
earlier than the formation of black
hole horizon.\\

Further, if $T_{1}$ and $T_{2}$ be the time differences between
the formation of cosmological horizon and singularity and the
formation of black hole horizon and singularity respectively then

\begin{equation}
T_{i}=\frac{2}{\sqrt{3\Lambda}}~sinh^{-1}\left[\sqrt{\frac{\Lambda}{3F}}~R_{i}^{3/2}
\right],~~i=1,2.
\end{equation}

A straight forward calculation shows

\begin{equation}
\frac{dT_{1}}{dF}<0~~
\text{and}~~\frac{dT_{2}}{dF}>0,~~\frac{dT}{dF}<0,~~T=T_{1}-T_{2}.
\end{equation}

Thus the time difference between the formation of singularity and
cosmological horizon decreases with $F$ increases while the time
difference between the formation of singularity and black hole
horizon increases with $F$. As $F$ is related to the mass of the
collapsing system so for more massive quasi-spherical model, the
time of formation of singularity and cosmological horizon become
close to each other while the time difference between the
formation of black hole horizon and that of cosmological horizon
becomes smaller.\\

\section{\normalsize\bf{Conclusion}}

In this paper, the collapse of a quasi-spherical star is
considered where the exterior geometry corresponds to
Schwarzschild-de-Sitter space-time. The junction conditions on
the boundary show a energy conservation equation on it.\\

Due to the presence of the cosmological constant $\Lambda$, the
Newtonian force is given by (see equation (23)) [24]
$$
P(R)=-\frac{M}{R^{2}}+\frac{\Lambda}{3}~R
$$

For collapsing process the force should be attractive in nature
and as a result $R$ should always be less than
$\left(\frac{3M}{\Lambda}\right)^{1/3}$. Further, the rate of
collapse has the expression

\begin{equation}
\ddot{R}=-\frac{M}{2R^{2}}+\frac{\Lambda}{3}~R
\end{equation}

which shows that the presence of $\Lambda$-term slows down the
collapsing process and hence influences the time difference
between the formation of the apparent horizon and the
singularity. \\

As the presence of a cosmological constant (dark energy) induces a
potential barrier to the equation of motion so particles with a
small velocity are unable to reach the central object. This ideas
can be used astrophysically for a particle orbiting a black hole,
which contains dark energy and an estimation of minimum velocity
can be done for which the particle enters inside the black hole.
Consequently, the amount of dark energy in the black hole can be
calculated.\\

Lastly, due to the presence of the cosmological constant, there
are two physical horizons $-$ the black hole horizon and the
cosmological horizon. Further, for more massive collapsing system,
the time of formation of the two horizons become very close to
each other. Moreover, asymptotic flatness of the space-time is
violated due to the presence of the cosmological
constant.\\

{\bf Acknowledgement:}\\\\
One of the authors (SC) is thankful to CSIR, Govt. of India for
providing a research project No. 25(0141)/05/EMR-II . Also the
authors thank to IUCAA for warm hospitality where the major part
of the work was done. \\

{\bf References:}\\
\\
$[1]$  J. R. Oppenhiemer and H. Snyder, {\it Phys. Rev.} {\bf 56} 455 (1939).\\
$[2]$  C. W. Misner and D. Sharp, {\it Phys. Rev.} {\bf 136} b571
(1964).\\
$[3]$  P. C. Vaidya, {\it Proc. Indian Acad. Sci. A} {\bf 33} 264
(1951).\\
$[4]$  N. O. Santos, {\it Phys. Lett. A} {\bf 106} 296 (1984).\\
$[5]$  N. O. Santos, {\it Mon. Not. R. Astr. Soc.} {\bf 216} 403
(1985).\\
$[6]$  A. K. G. de Oliveira, N. O. Santos and C. A. Kolassis,
{\it Mon. Not. R. Astr. Soc.} {\bf 216} 1001
(1985).\\
$[7]$  A. K. G. de Oliveira, J. A. de F. Pacheco and N. O. Santos,
{\it Mon. Not. R. Astr. Soc.} {\bf 220} 405
(1986).\\
$[8]$  A. K. G. de Oliveira and N. O. Santos, {\it Astrophys. J.}
{\bf 312} 640 (1987).\\
$[9]$  A. K. G. de Oliveira, C. A. Kolassis and N. O. Santos, {\it
Mon. Not. R. Astr. Soc.} {\bf 231} 1011 (1988).\\
$[10]$  S. G. Ghosh and D. W. Deshkar, {\it Int. J. Mod. Phys. D}
{\bf 12} 317 (2003).\\
$[11]$  S. G. Ghosh and D. W. Deshkar, {\it Gravitation and
Cosmology} {\bf 6} 1 (2000).\\
$[12]$  M. Cissoko, J. Fabris, J. Gariel, G. L. Denmat and N. O.
Santos, {\it gr-qc}/9809057; S. M. C. V. Goncalves,
{\it Class. Quantum Grav.} {\bf 18} 4517-4530 (2001).\\
$[13]$ D. F. Mota and C. Van de Bruc, {\it Astron. Astrophys.}
{\bf 421} 71 (2004).\\
$[14]$ R. -G. Cai and A. Wang, {\it gr-qc}/0505136.\\
$[15]$ O. Lahav, P. B. Lilje, J. R. Primack and M. J. Rees, {\it
Mon. Not. R. Astron. Soc.} {\bf 128} (1991).\\
$[16]$ A. Balaguera-Antolinez, C. G. Boehmer and M. Nowakowski,
{\it Class. Quantum Grav.} {\bf 23} 485 (2006).\\
$[17]$ S. Chakraborty and U. Debnath, {Int. J. Mod. Phys. D} {\bf 13} 1085 (2004); {\it gr-qc}/0304072.\\
$[18]$ P. Szekeres, {\it Phys. Rev. D} {\bf 12} 2941 (1975).\\
$[19]$ U. Debnath, S. Chakraborty and J. D. Barrow, {\it Gen. Rel.
Grav.}, {\bf 36} 231 (2004); {\it gr-qc}/0305075.\\
$[20]$ U. Debnath, S. Nath and S. Chakraborty, {\it Gen. Rel. Grav.} {\bf 37} 215 (2005 ).\\
$[21]$ P. Szekeres, {\it Commun. Math. Phys.} {\bf 41} 55 (1975).\\
$[22]$ W. Israel, {\it Nuovo Cimento} {\bf 44B} 1 (1966).\\
$[23]$ L. P. Eisenhart, {\it Riemannian Geometry}, p. 146 (1949),
Princeton.\\
$[24]$ S. Engineer, N. Kanekar and T. Padmanabhan, {\it Mon. Not.
R. Astron. Soc.} {\bf 314} 279 (2000).\\

\end{document}